\documentclass[journal,10pt,twoside]{IEEEtran}
\usepackage[T1]{fontenc}
\usepackage{lmodern}
\usepackage{microtype}
\usepackage{amsmath,amssymb,mathtools}
\usepackage{booktabs,tabularx,multirow}
\usepackage{graphicx}
\usepackage{caption}
\usepackage{algorithm,algorithmic}
\usepackage[numbers,sort&compress]{natbib}
\usepackage{tikz}
\usetikzlibrary{arrows.meta,positioning,fit,backgrounds}
\usepackage[hidelinks]{hyperref}
\usepackage{xcolor}

\title{RISK-KV: Quality-Aware Selective KV Cache Reconstruction for Distributed LLM Inference}
\author{Wang Yuekai$^{1}$\\
\small $^{1}$Harbin Institute of Technology\\
\small Email: \texttt{york.wang.ict@gmail.com}\\
}
\IEEEtitleabstractindextext{\begin{abstract}
\bfseries Request migration and interruption recovery in distributed large language model (LLM) inference rely on historical key--value (KV) caches. Retaining or transferring a complete cache incurs substantial state overhead, whereas replaying historical tokens requires additional computation. This paper proposes RISK-KV, a quality-aware method for selective historical KV reconstruction in pretrained models. RISK-KV uses exact KV states on both sides of a target layer and cross-layer ridge regression to predict the missing layer's attention-projection input; it then folds in the original model's KV projection matrices to reduce predictor storage. Reconstruction risk is measured by the change in negative log-likelihood (NLL) caused by actual reconstruction, and low-overhead topology search with quality gating selects the layer groups to reconstruct. Independent tests on three language models show that representative recovery configurations omit 18.2\%--33.3\% of historical KV, with perplexity changes below 0.5\%. In a local request-recovery experiment with a 512-token history, RISK-KV achieves a 1.51--3.31$\times$ speedup over full Token Replay. These results indicate that selective KV reconstruction can reduce historical-cache overhead with small changes in predictive quality and improve request-recovery efficiency under the evaluated deployment conditions.
\end{abstract}\begin{IEEEkeywords}
\normalfont KV cache, LLM inference, cache reconstruction, grouped-query attention, state migration, Transformer
\end{IEEEkeywords}}
\begin{document}
\maketitle
\IEEEdisplaynontitleabstractindextext
\IEEEpeerreviewmaketitle

\begin{figure*}[!t]
    \centering
    \includegraphics[width=0.95\textwidth]{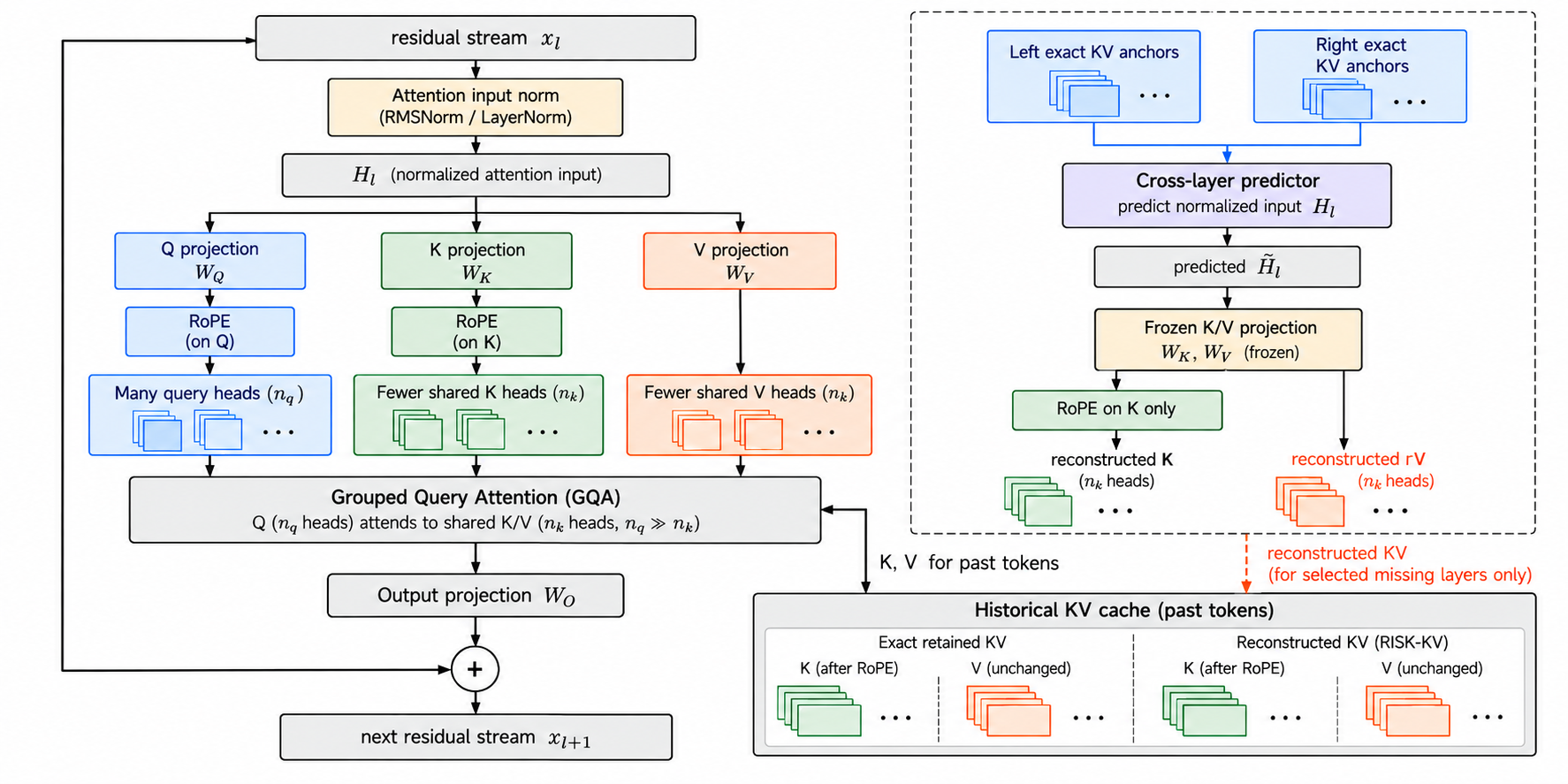}
    \caption{GQA attention module and cross-layer KV reconstruction in RISK-KV. The left side shows normalization, Q/K/V projections, RoPE, and the attention residual connection. The right side illustrates reconstruction of a target layer's historical keys and values from exact KV anchors on both sides. In the online implementation, hidden-representation prediction and KV projection are folded into an equivalent mapping.}
    \label{fig:gqa_riskkv}
\end{figure*}

\section{Introduction}
Large language models (LLMs) are increasingly deployed in distributed and edge-computing environments to support interactive and context-intensive applications. Autoregressive inference relies on a key--value (KV) cache to store the attention states of historical tokens and avoid redundant computation during subsequent decoding. However, KV-cache size grows with context length, model depth, and the number of concurrent requests, resulting in substantial memory and data-movement costs. vLLM demonstrates the importance of efficient KV management for LLM serving, while grouped-query attention (GQA) reduces cache size by sharing key/value heads \cite{kwon2023vllm,ainslie2023gqa}. In distributed inference, request migration and interruption recovery also involve transferring and reloading historical states. DistServe and Splitwise illustrate the potential of organizing inference computation across devices and highlight the effect of state management on system efficiency \cite{zhong2024distserve,patel2023splitwise}. In bandwidth-constrained edge environments, transferring the complete historical KV cache can be costly; discarding it instead requires recomputing the historical prefix, which also increases recovery latency.

Existing research reduces KV-cache overhead from several directions. H$_2$O and SnapKV use attention features to select historical tokens for retention, while KVQuant reduces cache storage through low-precision representations \cite{zhang2023h2o,li2024snapkv,hooper2024kvquant}. Another line of work exploits redundancy across Transformer layers: MiniCache compresses the cache by merging states across layers, whereas YOCO reuses KV states through its architectural design \cite{liu2024minicache,sun2024yoco}. For state transfer and recovery, CacheGen studies KV compression and streaming, while HCache uses intermediate activations to restore state and balance computation against I/O costs \cite{liu2024cachegen,gao2025hcache}. Krul dynamically selects cross-layer KV-sharing strategies based on attention similarity between layers; EdgeFlow studies KV migration and recovery in geographically distributed edge environments \cite{wen2025krul,zhang2026edgeflow}. These works show that exploiting cache redundancy and balancing storage, communication, and computation are important ways to improve inference efficiency.

A key question remains for reconstructing omitted historical KV without modifying a pretrained model: which layers are suitable for reconstruction, and which must retain exact KV to reduce cache overhead while controlling degradation in predictive quality? Existing methods typically make selection decisions using signals such as attention importance, tensor similarity, numerical reconstruction error, or recovery cost, but these signals do not necessarily reflect how reconstructed KV affects language-model predictions \cite{zhang2023h2o,li2024snapkv,liu2024minicache,lin2026fusedkv,liu2024cachegen,gao2025hcache}. Large numerical errors in KV may cause only a small increase in predictive loss; conversely, reconstruction errors in sensitive layers may propagate through subsequent Transformer layers and substantially change model outputs. Moreover, the effect of reconstructing several layer groups together need not equal the sum of their individual reconstruction losses. Therefore, maximizing the number of omitted layers alone may lead to an unfavorable trade-off between cache savings and model quality.

A key observation is that actual KV-reconstruction risk varies substantially across models and layers. With the remaining layers' KV kept exact, historical KV is replaced one layer at a time, and model degradation is measured by the change in negative log-likelihood (NLL) under teacher-forced inference. For example, on TinyLlama-1.1B, reconstructing L2 alone increases mean NLL by only about 0.00010, whereas reconstructing the adjacent layer L3 increases it by about 0.09336. Qwen2.5-0.5B also exhibits a marked risk difference between adjacent layers. Further experiments show that even when artificial Gaussian perturbations are matched in magnitude to actual reconstruction errors, the resulting rankings of layer sensitivity can differ. This indicates that recovery strategies should be calibrated to the target model and the specific reconstruction mechanism, rather than determined solely by layer position, reconstruction error, or generic perturbation sensitivity.

Based on these observations, this paper introduces RISK-KV, a quality-aware framework for selective historical KV reconstruction in distributed LLM inference. As illustrated in Fig.~\ref{fig:gqa_riskkv}, in a Transformer with grouped-query attention (GQA), historical KV is obtained by applying the key and value projections to attention-normalized hidden representations, with rotary positional embedding (RoPE) additionally applied to keys. RISK-KV uses exact historical KV retained on both sides of a target layer as cross-layer anchors. A lightweight ridge regressor predicts the target layer's normalized attention input, and the frozen key and value projection matrices of the original model generate the missing cache. Because both the predicted mapping and KV projections are affine transformations, they can be algebraically folded so that the online stage generates reconstructed KV directly from the anchor states, reducing additional parameter storage and intermediate-representation computation. This process neither modifies pretrained-model parameters nor requires storing complete historical hidden states.

To identify layer groups suitable for reconstruction, RISK-KV defines functional risk as the mean NLL change caused by replacing actual KV. The system first measures the independent recovery risk of each candidate group and ranks the groups accordingly. It then progressively forms joint recovery configurations along this ranking and remeasures the actual predictive loss for each configuration, rather than simply summing the individual group risks. For $G$ candidate groups, the method evaluates only $2G$ distinct recovery configurations, compared with $2^G$ for exhaustive search. RISK-KV also uses a conservative quality gate based on bootstrap confidence bounds to select a configuration that satisfies the empirical quality budget while saving substantial historical KV, falling back to exact KV when necessary. Because the search is restricted to a single ordered path, it does not guarantee a global optimum, but it reduces offline configuration-evaluation cost.

RISK-KV is evaluated on three pretrained models: Qwen2.5-0.5B-Instruct, Qwen2.5-1.5B-Instruct, and TinyLlama-1.1B-Chat-v1.0. On independent test data, representative recovery configurations omit 18.2\%--33.3\% of historical KV, with changes in perplexity relative to the full exact-KV baseline below 0.5\%. The quality-gating experiments further show that more conservative selection can reduce observed budget violations in independent evaluations, typically at the cost of some cache savings. In local request-recovery experiments with a 512-token history, RISK-KV achieves a 1.51--3.31$\times$ speedup over full historical Token Replay. The current latency experiments do not cover real wireless cross-node migration, and the complete KV cache is still materialized during subsequent decoding.

\paragraph*{Contributions.} This paper makes the following contributions:
\begin{enumerate}
\item A method is proposed for reconstructing historical KV caches from cross-layer information. To address the loss or unavailability of some historical KV states during request recovery in distributed LLM inference, the method uses real KV retained in other layers to reconstruct missing keys and values in target layers. It requires neither changes to pretrained-model parameters nor storage of complete historical hidden states or recomputation of the entire historical prefix. By folding the model's KV projection matrices into the predictor, it reduces additional reconstruction storage and provides an implementation path for reducing persistent storage and transfer of historical KV.
\item A functional-risk assessment and adaptive selection method is proposed for KV reconstruction across models and layers. The results show substantial variation in the tolerance of different models and Transformer layers to historical KV-reconstruction errors; KV numerical error or random-perturbation sensitivity alone cannot reliably identify suitable layers. Accordingly, the increase in language-model NLL caused by actual KV reconstruction is used as a measure of model-degradation risk. A small number of singleton recovery experiments assess candidate-layer risks, and low-overhead topology search under a quality constraint, together with conservative quality gating, automatically determines which historical KV states require exact preservation and which can be reconstructed, enabling a controlled trade-off between predictive quality and cache savings.
\end{enumerate}
\section{Related Work}
\subsection{KV Cache Management and Compression}
KV caching avoids recomputing attention states for historical tokens, but its memory footprint grows with context length and request concurrency. Existing methods address this cost in different ways. PagedAttention organizes KV storage in blocks to improve memory allocation efficiency and reduce fragmentation in LLM serving \cite{kwon2023vllm}. Grouped-query attention (GQA) changes the attention architecture by letting multiple query heads share key/value heads, thereby reducing the number of KV heads that must be stored \cite{ainslie2023gqa}. Token-selection methods directly reduce the number of retained positions: H$_2$O dynamically preserves recent tokens and a small set of attention heavy hitters, while SnapKV selects salient positions based on attention observed near the end of the prompt \cite{zhang2023h2o,li2024snapkv}. KVQuant uses low-bit quantization to reduce the storage cost of KV representations \cite{hooper2024kvquant}. These methods therefore act on memory allocation, KV head count, token positions, or numerical precision.

\subsection{Cross-Layer KV Sharing and State Reconstruction}
Another line of work exploits dependencies across Transformer depth. MiniCache merges KV states from adjacent layers based on their similarity \cite{liu2024minicache}. FusedKV constructs top-layer KV caches by learning to fuse informative KV states from lower layers \cite{lin2026fusedkv}. Krul dynamically selects cross-layer sharing strategies based on attention similarity between layer pairs and combines compressed-cache loading with recomputation to restore state \cite{wen2025krul}. These methods explore different mechanisms for cross-layer sharing and restoration. Layer-wise functional risk under a particular reconstruction mechanism cannot be equated directly with attention similarity or tensor similarity.

Systems for transferring and restoring request state address complementary costs. CacheGen encodes KV caches as compact bitstreams and streams them to reduce context-loading overhead, adapting its compression scheme to changing bandwidth \cite{liu2024cachegen}. HCache restores model state from stored intermediate activations and schedules computation and I/O to balance the use of different resources during restoration \cite{gao2025hcache}. RISK-KV targets pretrained models. It uses exact historical KV retained on both sides of a target layer as cross-layer information anchors, then learns a mapping to reconstruct the normalized hidden state input to the target layer's attention module. The pretrained model's original KV projection matrices then map this reconstructed hidden representation to the missing historical keys and values, combining cross-layer state correlations with the model's existing projection structure. RISK-KV further measures each layer's functional risk using the NLL change caused by actual KV reconstruction, and selects which historical KV states to preserve exactly or reconstruct under a given quality budget. This approach trades cache savings against predictive quality.

\section{The RISK-KV Framework}
\label{sec}

\subsection{System Architecture and KV Cache Model}
\label{subsec:system_model}

RISK-KV addresses historical KV management and request state restoration in wireless distributed LLM inference. Consider a pretrained autoregressive model with $L$ Transformer decoder layers. Its parameters remain frozen throughout KV reconstruction. During inference, each layer stores the keys and values of historical tokens to avoid recomputing their attention states during subsequent decoding. Although grouped-query attention (GQA) reduces the number of KV heads, each Transformer layer still maintains its own historical KV cache \cite{ainslie2023gqa,kwon2023vllm}.

For a history of $T$ tokens, $n_{\mathrm{kv}}$ KV heads per layer, head dimension $d_h$, and $q$ bytes per cache element, the full historical KV cache occupies
\begin{equation}
M_{\mathrm{KV}}=2LTn_{\mathrm{kv}}d_hq.
\label{eq:kv_cache_size}
\end{equation}

\begin{figure*}[t]
    \centering
    \includegraphics[width=\textwidth]{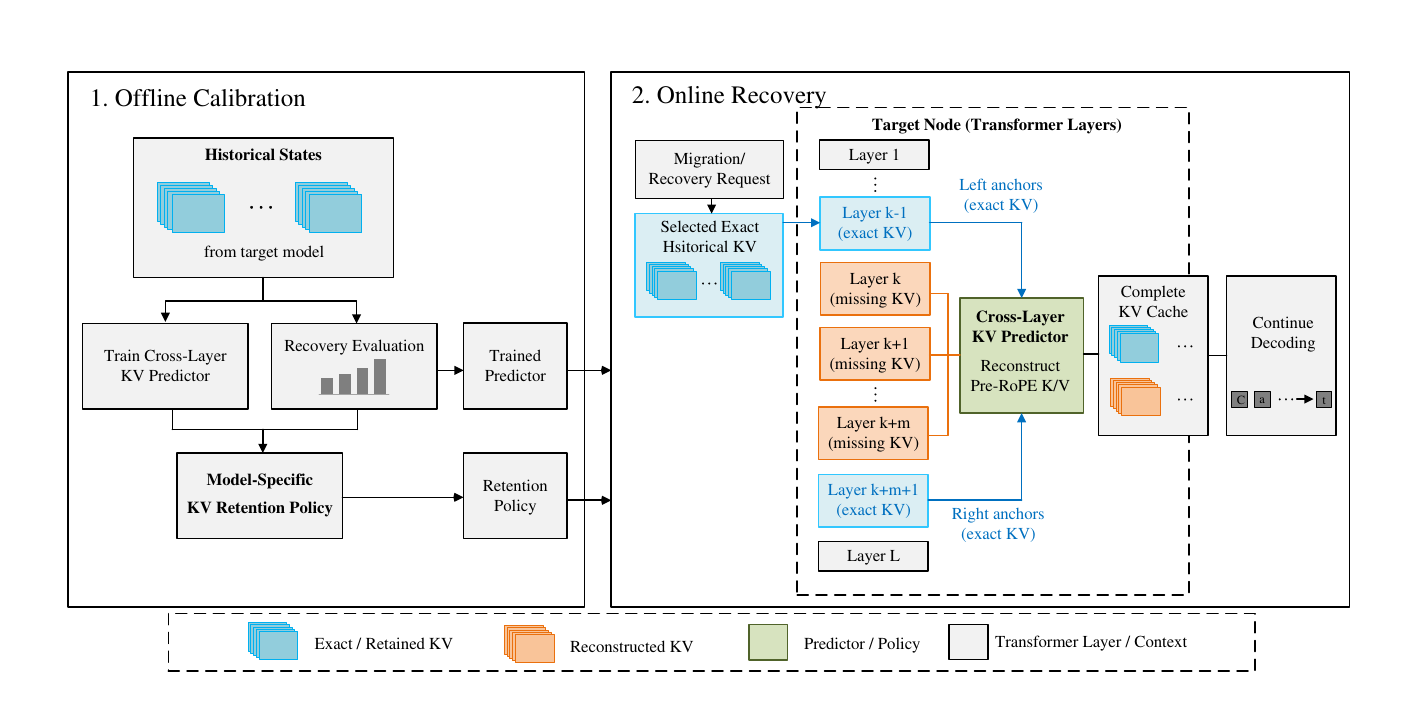}
    \caption{Offline calibration and online recovery in RISK-KV.}
    \label{fig:risk-kv-overview}
\end{figure*}

As shown in Fig.~\ref{fig:risk-kv-overview}, RISK-KV combines offline calibration with online recovery. Offline, historical states from the target model train a cross-layer KV predictor; experiments that actually replace KV determine layer-specific functional risk and produce a model-specific retention policy. When a request is migrated or restored online, the system retains or transfers exact historical KV only for selected layers and reconstructs the other layers with the deployed predictor.

For each candidate recovery group, RISK-KV places exact KV anchors on both sides along Transformer depth. Let $G_g$ be group $g$, with left and right anchor sets $\mathcal A_g^-$ and $\mathcal A_g^+$. To reconstruct a target layer's KV for one historical token, the predictor uses exact anchor KV at the same token position. Anchor layers retain their complete, exact histories and are never reconstructed. The procedure neither changes base-model parameters nor requires a complete historical hidden-state checkpoint or replay of the entire prefix.

\subsection{Cross-Layer KV Reconstruction from Two-Sided Anchors}
\label{subsec:cross_layer_reconstruction}

RISK-KV uses exact historical KV on both sides of a target layer to predict the normalized hidden input to that layer's attention projections. The pretrained model's original key and value projections then produce its missing KV.

For layer $\ell$, let $H_\ell\in\mathbb R^{T\times d}$ be the hidden representation after attention normalization and before the key/value projections. Its KV projection before rotary position encoding is
\begin{equation}
Y_\ell=H_\ell W_\ell+\mathbf{1}_T b_\ell^\top,
\label{eq:kv_projection}
\end{equation}
where $Y_\ell=[K_\ell^{\mathrm{pre}},V_\ell]$, $W_\ell=[W_\ell^K,W_\ell^V]$, $b_\ell$ is the frozen projection bias, and $\mathbf{1}_T$ is the all-ones vector of length $T$. For a model without a projection bias, $b_\ell=\mathbf0$.

For candidate group $G_g$, the method builds cross-layer features from its exact anchors. Keys in a standard KV cache have generally undergone rotary position encoding (RoPE). When the anchor keys come from a post-RoPE cache, inverse RoPE at the original token positions first returns them to pre-RoPE projection coordinates; values need no rotation. The pre-RoPE keys and values from both sides are then concatenated into the input features $X_g$.

Offline, ridge regression learns a mapping from these anchors to the complete attention-projection inputs of the target layers. Let $\mathcal T=G_g$ be a two-layer target group and $\mathcal A=\mathcal A_g^-\cup\mathcal A_g^+$ its exact anchors. Concatenate their pre-RoPE keys and values as input, and the normalized hidden projection inputs of the target layers as labels:
\begin{equation}
\begin{aligned}
X_{\mathcal A}&=\operatorname{Concat}_{j\in\mathcal A}[K_j^{\mathrm{pre}},V_j],\\
H_{\mathcal T}&=\operatorname{Concat}_{\ell\in\mathcal T}H_\ell.
\end{aligned}
\label{eq:ridge_features}
\end{equation}

Training standardizes the input features, centers only the target, and fits ridge regression with an intercept. After converting the fitted coefficients back to the original input coordinates, the prediction is
\begin{equation}
\hat H_{\mathcal T}=X_{\mathcal A}A_{\mathcal T}+\mathbf{1}_T b_{\mathcal T}^{\top},
\label{eq:group_hidden_recovery}
\end{equation}
where $A_{\mathcal T}$ and $b_{\mathcal T}$ are the predictor weight and bias. The predictor estimates each target layer's complete normalized attention input, rather than a residual relative to another hidden state. No historical hidden state or residual must therefore be retained as an additional input.

Form a block-diagonal matrix from the original frozen KV projections of the target layers, $P_{\mathcal T}=\operatorname{blkdiag}(W_\ell)_{\ell\in\mathcal T}$, and concatenate their projection biases as $c_{\mathcal T}=\operatorname{Concat}_{\ell\in\mathcal T}b_\ell$. The predicted pre-RoPE KV is
\begin{equation}
\hat Z_{\mathcal T}^{\mathrm{pre}}=\hat H_{\mathcal T}P_{\mathcal T}+\mathbf{1}_T c_{\mathcal T}^{\top},
\label{eq:group_kv_projection}
\end{equation}
which contains the predicted pre-RoPE keys and values for every target layer in the group.

Both hidden-state prediction and KV projection are affine. They can therefore be folded algebraically into
\begin{equation}
\hat Z_{\mathcal T}^{\mathrm{pre}}=X_{\mathcal A}B_{\mathcal T}+\mathbf{1}_T d_{\mathcal T}^{\top},
\label{eq:folded_kv_recovery}
\end{equation}
where
\begin{equation}
\begin{aligned}
B_{\mathcal T}&=A_{\mathcal T}P_{\mathcal T},\\
d_{\mathcal T}^{\top}&=b_{\mathcal T}^{\top}P_{\mathcal T}+c_{\mathcal T}^{\top}.
\end{aligned}
\label{eq:folded_kv_parameters}
\end{equation}
The folded predictor generates pre-RoPE KV directly from exact anchors without explicitly materializing the intermediate hidden representation. At historical position $t$, RoPE is reapplied only to the predicted key:
\begin{equation}
\hat K_{\ell,t}^{\mathrm{cache}}=\operatorname{RoPE}_t(\hat K_{\ell,t}^{\mathrm{pre}}),\qquad
\hat V_{\ell,t}^{\mathrm{cache}}=\hat V_{\ell,t}.
\label{eq:recovered_kv_cache}
\end{equation}
Folding is algebraically equivalent to explicit hidden-state prediction followed by KV projection in exact arithmetic; finite-precision paths can differ slightly. Online recovery uses only the retained exact anchors and the offline-trained predictor. True target hidden states serve as offline labels only, and neither the omitted exact KV nor the target hidden states are read online.

\subsection{Quality-Constrained KV Recovery Selection}
\label{subsec:risk_aware_selection}

Although two-sided anchors make reconstruction possible, its effect on predictive quality varies substantially by layer. Some layers tolerate numerical KV reconstruction error with little effect on predictions, whereas errors in others cause substantial loss. RISK-KV therefore measures the loss caused by actual KV replacement and selects recovery groups under a quality budget instead of reconstructing all candidates indiscriminately.

Let $\mathcal G=\{G_1,\ldots,G_G\}$ be the candidate groups. A recovery topology $S\subseteq\mathcal G$ specifies groups that use predicted KV; all other layers use exact KV. $S=\varnothing$ denotes the full exact cache. For calibration request $r$, $\ell_r(S)$ and $\ell_r(\varnothing)$ are mean negative log-likelihoods (NLLs) on the same target tokens. The empirical functional risk is
\begin{equation}
\hat D(S)=\frac{1}{N}\sum_{r=1}^{N}\left[\ell_r(S)-\ell_r(\varnothing)\right],
\label{eq:empirical_risk}
\end{equation}
where $N$ is the number of calibration requests. Unlike numerical key/value error, $\hat D(S)$ directly measures the effect of reconstructed KV on predictive loss.

RISK-KV first recovers each group alone, keeping all other layers exact, and obtains singleton risk $\rho_g=\hat D(\{G_g\})$. It ranks groups by increasing risk per savable byte, $\rho_g/M_g$, where $M_g=\sum_{\ell\in G_g}M_\ell$. The groups in our experiments are equally sized, so this is the same as ranking by singleton risk. If $\pi=(\pi_1,\ldots,\pi_G)$ is the resulting order, the combined candidates are
\begin{equation}
S_k=\{G_{\pi_1},\ldots,G_{\pi_k}\},\quad k=1,\ldots,G.
\label{eq:ordered_topology}
\end{equation}
Because jointly reconstructed groups can have non-additive effects, each combined topology is evaluated directly; singleton losses are not summed. Counting the exact baseline, $G$ singletons, and $G-1$ additional prefixes gives $2G$ distinct evaluations, versus $2^G$ for exhaustive enumeration. Restricting the search to this risk-ordered path reduces offline evaluations but offers no general global-optimality guarantee.

To account for finite calibration samples, RISK-KV bootstraps the paired request losses and computes a one-sided 90\% confidence upper bound $U_{0.90}(S)$ for mean degradation. Given an allowed mean NLL increase $\epsilon$, it accepts only configurations satisfying
\begin{equation}
U_{0.90}(S)\leq\epsilon.
\label{eq:upper_confidence_gate}
\end{equation}
Among evaluated candidates $\mathcal C$, it then chooses the feasible topology omitting the most historical KV. Algorithm~\ref{alg} summarizes the offline procedure:
\begin{equation}
\begin{aligned}
S^*=\arg\max_{S\in\mathcal C}\quad &M_{\mathrm{saved}}(S),\\
\mathrm{s.t.}\quad &U_{0.90}(S)\leq\epsilon.
\end{aligned}
\label{eq:topology_optimization}
\end{equation}
$M_{\mathrm{saved}}(S)$ is the number of omitted historical KV bytes. If no nonempty candidate passes the gate, recovery falls back to the full exact cache. All risk probing, topology evaluation, and gating occur offline; online request recovery applies the selected topology without remeasuring predictive loss.

\begin{algorithm}[t]
\caption{Quality-constrained KV recovery topology selection}
\label{alg}
\begin{algorithmic}[1]
\REQUIRE Candidate groups $\mathcal G$, calibration set $\mathcal D_{\mathrm{cal}}$, quality budget $\epsilon$
\ENSURE Recovery topology $S^*$
\STATE Evaluate the full exact-KV baseline
\STATE $\mathcal C\leftarrow\varnothing$
\FOR{each candidate group $G_g\in\mathcal G$}
\STATE Evaluate singleton risk $\rho_g=\hat D(\{G_g\})$
\STATE $\mathcal C\leftarrow\mathcal C\cup\{\{G_g\}\}$
\ENDFOR
\STATE Sort groups by increasing $\rho_g/M_g$ to obtain $\pi$
\FOR{$k=2,\ldots,G$}
\STATE Form $S_k=\{G_{\pi_1},\ldots,G_{\pi_k}\}$
\STATE Evaluate joint risk $\hat D(S_k)$
\STATE $\mathcal C\leftarrow\mathcal C\cup\{S_k\}$
\ENDFOR
\FOR{each $S\in\mathcal C$}
\STATE Compute upper confidence bound $U_{0.90}(S)$
\ENDFOR
\STATE $\mathcal F\leftarrow\{S\in\mathcal C:U_{0.90}(S)\leq\epsilon\}$
\STATE $S^*\leftarrow\varnothing$ if $\mathcal F=\varnothing$; otherwise $S^*\leftarrow\arg\max_{S\in\mathcal F}M_{\mathrm{saved}}(S)$
\RETURN $S^*$
\end{algorithmic}
\end{algorithm}

\section{Storage and Recovery Cost}
\label{sec:cost_analysis}

Selective historical KV reconstruction reduces retained cache state but adds predictor storage and recovery computation. We analyze these costs for cross-node migration and local request restoration.

\subsection{Storage and Communication}

Let $S$ be the selected topology and $\mathcal L(S)$ its reconstructed target layers. The omitted historical KV occupies
\begin{equation}
M_{\mathrm{saved}}(S)=\sum_{\ell\in\mathcal L(S)}M_\ell
\label{eq:cost_kv_saved}
\end{equation}
where $M_\ell$ is layer $\ell$'s historical KV size.

RISK-KV also stores its folded predictors. If predictor $g$ has input and output dimensions $d_{x,g}$ and $d_{y,g}$ and each weight uses $q_w$ bytes, the selected predictors occupy
\begin{equation}
M_{\mathrm{pred}}(S)=q_w\sum_{G_g\in S}d_{y,g}(d_{x,g}+1).
\label{eq:cost_predictor_size}
\end{equation}
Predictor parameters are shared by requests using the same model. For $N$ requests with the same topology, if omitted KV is actually evicted and the predictor resides at the same storage tier, net savings relative to full KV are
\begin{equation}
\Delta M_{\mathrm{store}}=NM_{\mathrm{saved}}(S)-M_{\mathrm{pred}}(S)-M_{\mathrm{aux}},
\label{eq:cost_net_storage}
\end{equation}
where $M_{\mathrm{aux}}$ denotes additional resident auxiliary state.

For migration from server A to server B, transfer volumes for full KV and RISK-KV are
\begin{equation}
\begin{aligned}
M_{\mathrm{tx}}^{\mathrm{full}}&=M_{\mathrm{KV}}+M_{\mathrm{meta}},\\
M_{\mathrm{tx}}^{\mathrm{RISK}}(S)&=M_{\mathrm{tx}}^{\mathrm{full}}-M_{\mathrm{saved}}(S)\\
&\quad+M_{\mathrm{extra}}+I_{\mathrm{init}}M_{\mathrm{pred}}(S).
\end{aligned}
\label{eq:cost_transfer}
\end{equation}
Here $M_{\mathrm{meta}}$ is common cache metadata, $M_{\mathrm{extra}}$ accounts for additional recovery information and anchor communication, and $I_{\mathrm{init}}\in\{0,1\}$ indicates whether predictors must be transferred. When they are already deployed at the destination, $I_{\mathrm{init}}=0$. Net communication savings depend on whether omitted KV exceeds these additional transfer costs.

\subsection{Recovery Latency}

Let $R_{\mathrm{eff}}$ be the effective inter-server rate in bits per second. For serial cross-node migration, the two recovery times can be written as
\begin{equation}
t_{\mathrm{mig}}^{\mathrm{full}}=\frac{8M_{\mathrm{tx}}^{\mathrm{full}}}{R_{\mathrm{eff}}}+t_{\mathrm{ov}}^{\mathrm{full}},
\label{eq:cost_full_latency}
\end{equation}
\begin{equation}
\begin{aligned}
t_{\mathrm{mig}}^{\mathrm{RISK}}(S)={}&\frac{8M_{\mathrm{tx}}^{\mathrm{RISK}}(S)}{R_{\mathrm{eff}}}\\
&+t_{\mathrm{ov}}^{\mathrm{RISK}}+t_{\mathrm{rec}}(S).
\end{aligned}
\label{eq:cost_risk_latency}
\end{equation}
$t_{\mathrm{ov}}$ includes serialization, deserialization, cache assembly, and any necessary predictor loading. $t_{\mathrm{rec}}(S)$ covers anchor-feature processing, KV prediction, and RoPE. Their latency difference is $t_{\mathrm{mig}}^{\mathrm{full}}-t_{\mathrm{mig}}^{\mathrm{RISK}}(S)$.

Without migration, a suspended request can retain exact anchors and evict reconstructible historical KV. RISK-KV avoids replaying the entire prefix and reaches the next token in
\begin{equation}
\begin{aligned}
t_{\mathrm{resume}}^{\mathrm{RISK}}={}&t_{\mathrm{fetch}}^{\mathrm{KV}}+t_{\mathrm{rec}}(S)\\
&+t_{\mathrm{mat}}+t_{\mathrm{dec}}^{\mathrm{RISK}},
\end{aligned}
\label{eq:cost_resume_latency}
\end{equation}
where $t_{\mathrm{fetch}}^{\mathrm{KV}}$ reads retained KV, $t_{\mathrm{mat}}$ assembles the recovered cache, and $t_{\mathrm{dec}}^{\mathrm{RISK}}$ computes the next token. If all historical KV is evicted, token replay instead takes
\begin{equation}
t_{\mathrm{resume}}^{\mathrm{replay}}=t_{\mathrm{fetch}}^{\mathrm{token}}+t_{\mathrm{prefill}}(T)+t_{\mathrm{dec}}^{\mathrm{exact}},
\label{eq:cost_replay_latency}
\end{equation}
where the terms read historical token IDs and rerun the length-$T$ prefix to rebuild exact KV. Thus, a recovery benefit depends on whether saved transfer or recomputation time exceeds prediction and cache-assembly costs. Longer contexts may offer greater potential savings against replay, but end-to-end performance depends on hardware, storage, and network conditions. Our local latency measurements cover only 512-token histories.

\section{Performance and Analysis}
\label{sec:performance}

\subsection{Experimental Setup}
\label{subsec:experimental_setup}

We evaluate RISK-KV on three pretrained autoregressive language models: Qwen2.5-0.5B-Instruct, Qwen2.5-1.5B-Instruct, and TinyLlama-1.1B-Chat-v1.0, with 24, 28, and 22 Transformer layers, respectively. Base-model parameters remain frozen in every experiment; only the cross-layer KV recovery predictors are fitted on training data.

Experiments use WikiText-2 with a 512-token historical context. Predictive quality after recovery is evaluated by teacher forcing over the following 512 tokens. The three models have 5, 6, and 5 two-layer candidate recovery groups, respectively, with exact KV always retained at both anchors. Their full historical KV sizes are 6, 14, and 11 MiB.

RISK-KV predicts each target layer's complete normalized attention-projection input using ridge regression with an intercept, then generates reconstructed KV through the original frozen projections. Deployment-cost experiments additionally evaluate folded FP32 and FP16 predictors. Every recovery configuration is compared with a full exact-KV baseline.

We measure predictive quality using mean change in negative log-likelihood ($\Delta\mathrm{NLL}$) and perplexity (PPL). Omitted raw KV bytes and their fraction of the full cache measure potential cache savings. The statistical unit is a request: recovered and exact KV are compared on the same request, and confidence intervals are estimated by bootstrap. Predictive-quality experiments use three random seeds with eight evaluation requests per seed; the local recovery-latency experiment below uses eight validation requests per model.

The data splits serve distinct purposes. Training data fit the ridge predictors; validation data select recovery groups; an independent test split evaluates the final frozen recovery configurations. The quality-gate experiment also uses separate requests for risk calibration and outcome assessment to avoid reusing evaluation data in selection.

\subsection{Experimental Results}
\label{subsec:experimental_results}

\begin{figure*}[t]
    \centering
    \includegraphics[width=0.85\textwidth]{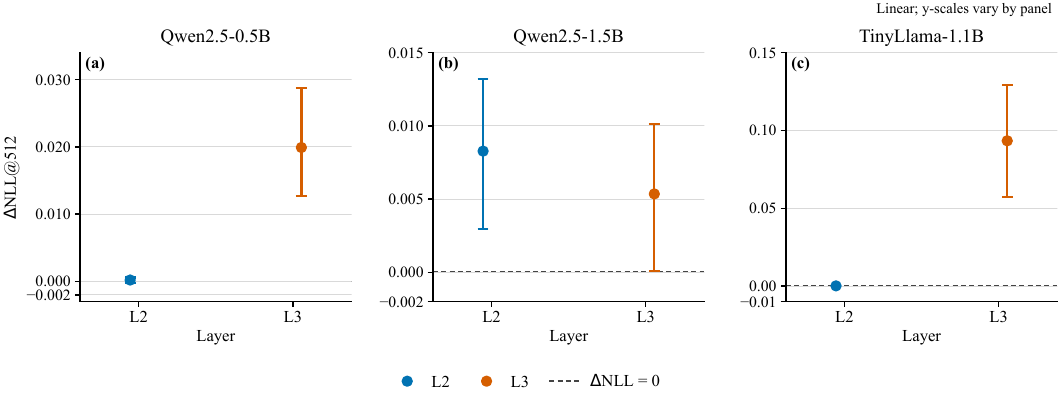}
    \caption{Functional risk from reconstructing a single layer in each model. Points show mean paired $\Delta$NLL@512 after reconstructing L2 or L3; error bars show paired 95\% bootstrap confidence intervals relative to the exact-KV baseline. Vertical scales differ across panels.}
    \label{fig:layer_risk}
\end{figure*}

We first examine how reconstructing different Transformer layers affects predictive quality. Figure~\ref{fig:layer_risk} compares the singleton recovery risks of adjacent layers L2 and L3 in all three models. Each validation result uses 24 requests (three seeds, eight per seed). Only the specified layer's historical KV is reconstructed; all other layers retain exact KV. Teacher forcing measures NLL change over the following 512 tokens.

L3 has a much higher reconstruction loss than L2 in Qwen2.5-0.5B and TinyLlama-1.1B, whereas Qwen2.5-1.5B shows the opposite ordering. Thus, even with the same reconstruction method, adjacent layers can differ greatly in their predictive effect, and the high-risk position changes across models. Layer position alone cannot determine which KV to reconstruct; selection must account for the loss caused by actual reconstruction.

\begin{figure}[t]
    \centering
    \includegraphics[width=\linewidth]{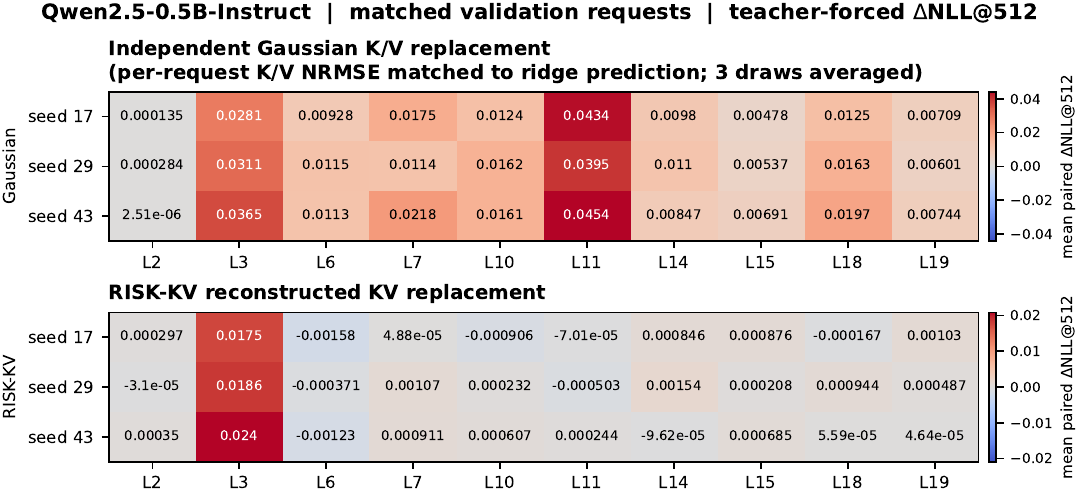}
    \caption{Per-layer paired $\Delta\mathrm{NLL}@512$ from independent Gaussian KV replacement and RISK-KV reconstructed KV replacement on Qwen2.5-0.5B-Instruct. Both panels use the same three seeds, training data, two-sided exact anchors, and eight validation requests per seed; all historical KV outside the target layer is exact. Gaussian K/V NRMSE is matched to the RISK-KV prediction per request and layer, with three independent draws averaged. Color ranges are scaled separately; compare annotated values rather than colors across panels.}
    \label{fig:gaussian_prediction_all_layers}
\end{figure}

Figure~\ref{fig:gaussian_prediction_all_layers} also compares independent Gaussian KV replacement with RISK-KV reconstruction across all ten deployable recovery layers of Qwen2.5-0.5B-Instruct. Both methods use the same three seeds, training data, two-sided exact anchors, eight validation requests per seed, and 512-token teacher-forced evaluation window. All historical KV outside the target layer remains exact. The Gaussian control matches the K/V normalized root mean square error (NRMSE) of the RISK-KV prediction separately for each request and layer, averaging three noise draws. It controls numerical error magnitude while retaining the difference between random and structured reconstruction-error directions. Gaussian replacement's largest mean loss is at L11 (0.04277); reconstructed KV's largest positive mean loss is at L3 (0.02002). Even at matched request-level K/V NRMSE, a sensitivity ranking based on artificial isotropic noise is unsuitable for choosing RISK-KV recovery layers.

We next examine the trade-off between predictive quality and cache savings. Table~\ref{tab:reconstruction_quality} gives representative results on independent test data. Each topology was selected on validation data and frozen before testing; no test request influenced selection. Each model has 24 paired test requests.
\begin{table*}[t]
\centering
\caption{Representative validation-frozen topologies on the H3 sealed test. $\Delta$NLL intervals are paired 95\% bootstrap confidence intervals. PPL summaries and paired request-level $\Delta$NLL use different aggregation orders, so $\Delta$NLL cannot be recovered from the ratio of the two displayed PPL values.}
\label{tab:reconstruction_quality}
\footnotesize
\begin{tabular}{l l c c c c l}
\toprule
Model & Recovered groups & KV saved & Exact PPL & RISK-KV PPL & $\Delta$NLL & 95\% CI \\
\midrule
Qwen2.5-0.5B & G2,G3,G4,G5 & 2.0 MiB (33.3\%) & 22.760 & 22.801 & 0.001622 & [$-0.000571$, 0.003962] \\
Qwen2.5-1.5B & G2,G3,G5,G6 & 4.0 MiB (28.6\%) & 15.205 & 15.183 & $-0.000207$ & [$-0.008627$, 0.006686] \\
TinyLlama-1.1B & G2,G5 & 2.0 MiB (18.2\%) & 12.681 & 12.722 & 0.003645 & [$-0.002192$, 0.008021] \\
\bottomrule
\end{tabular}
\end{table*}

\begin{figure*}[t]
    \centering
    \includegraphics[width=0.85\textwidth]{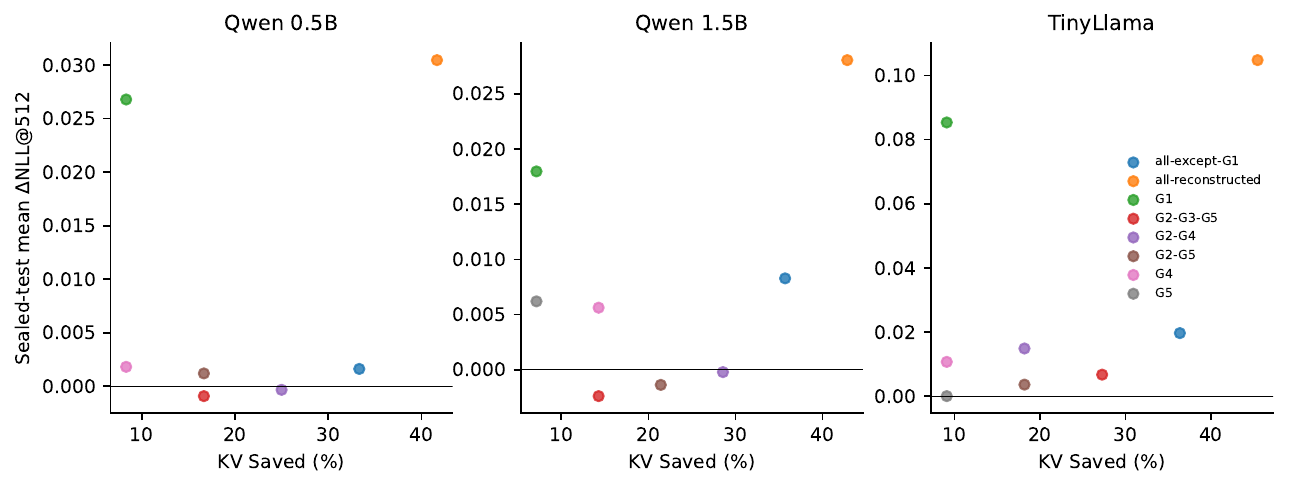}
    \caption{Historical KV saved versus predictive loss for different recovery topologies. Each point is a validation-preselected configuration evaluated on an independent test set.}
    \label{fig:sealed_frontier}
\end{figure*}

Some configurations have negative $\Delta$NLL: in the finite test sample, their mean predictive loss is slightly below the exact-KV baseline. This does not establish that approximate reconstruction improves quality in general, because KV replacement changes the predictive distribution and its effect can vary across requests. For example, Qwen2.5-1.5B has mean $\Delta$NLL $-0.000207$, but its paired 95\% interval includes zero; we therefore cannot conclude that quality improved.

Figure~\ref{fig:sealed_frontier} relates historical KV savings to $\Delta$NLL for other recovery configurations. Configurations with similar cache savings can have different predictive losses, and degradation becomes pronounced for some configurations as more layers are reconstructed. Maximizing cache reduction alone is therefore insufficient; the quality budget should be applied to measured recovery risk.

For $G$ candidate groups, RISK-KV measures singleton risk and then evaluates joint configurations along the risk order, requiring $2G$ topology evaluations. Five and six candidates require 10 and 12 evaluations, compared with 32 and 64 for exhaustive search. In some quality-budget settings the method matches exhaustive KV savings, but global optimality is not guaranteed. In an independent assessment at $\epsilon=0.005$, mean historical KV savings are 33.3\%, 33.3\%, and 21.2\% for Qwen2.5-0.5B, Qwen2.5-1.5B, and TinyLlama-1.1B, respectively.

Risk-ranked selection reduces retained historical KV, but cache savings do not ensure that every new request meets the quality budget. Requests differ in sensitivity: a configuration with low mean NLL increase on calibration data can still produce a large loss on a new request.

We therefore compare three quality gates. Mean Gate accepts a topology based only on mean calibration $\Delta$NLL. Upper-90 applies a more conservative one-sided 90\% upper confidence bound on mean loss. Stable-split additionally checks candidates on an independent confirmation set. Table~\ref{tab:quality_gate} reports TinyLlama-1.1B at $\epsilon=0.002$, the allowed mean NLL increase. The violation rate counts independent evaluation requests whose observed $\Delta$NLL exceeds $\epsilon$; mean KV saved measures the historical cache omitted by each strategy.
\begin{table}[t]
\centering
\caption{Predictive quality and KV savings of quality gates on TinyLlama-1.1B ($\epsilon=0.002$, 24 independent evaluation requests).}
\label{tab:quality_gate}
\footnotesize
\begin{tabular}{lcc}
\toprule
Gate & Budget violations & Mean KV saved \\
\midrule
Mean Gate & 41.7\% (10/24) & 1.667 MiB \\
Upper-90 & 8.3\% (2/24) & 0.667 MiB \\
Stable-split & 0\% (0/24) & 0.333 MiB \\
\bottomrule
\end{tabular}
\end{table}
In this sample, Upper-90 has eight fewer observed violating requests than Mean Gate, but also saves less KV. Stable-split has no observed violations, yet falls back to exact KV for two thirds of requests and saves still less. Exact fallback rates are 0\%, 33.3\%, and 66.7\%, respectively, illustrating the empirical trade-off between quality control and cache reduction. Whether omitted KV covers predictor cost also depends on per-group parameter storage and local recovery overhead.

To assess the additional storage cost, Table~\ref{tab:predictor_storage} compares the original ridge predictor with its folded form. Composing the recovery map with the model's original KV projections markedly reduces predictor storage, and FP16 halves it again. Values are average parameter sizes for one recovery group, rather than selected predictor-bank totals; deployment cost depends on the chosen groups and how many requests share the predictors.
\begin{table}[t]
\centering
\caption{Mean storage of one predictor group for each model (MiB).}
\label{tab:predictor_storage}
\footnotesize
\setlength{\tabcolsep}{4pt}
\begin{tabular}{lccc}
\toprule
Model & Original ridge & Folded FP32 & Folded FP16 \\
\midrule
Qwen2.5-0.5B & 7.007 & 2.002 & 1.001 \\
Qwen2.5-1.5B & 24.012 & 8.004 & 4.002 \\
TinyLlama-1.1B & 32.016 & 8.004 & 4.002 \\
\bottomrule
\end{tabular}
\end{table}

In a representative Qwen2.5-0.5B configuration, RISK-KV omits about 33.3\% of historical KV, but the selected Folded FP32 predictor parameters are about four times the KV saved by one request. For a batch of 24 requests, measured GPU prediction and full-cache assembly take 3.974 and 3.810 ms, respectively. At an assumed effective rate of 100 Mbps, the omitted 2 MiB per request corresponds to about 167.8 ms of pure data-transfer time. Thus, selective reconstruction has potential communication benefits if predictors are deployed in advance and shared across requests. This is a byte/rate calculation, not a measured wireless transfer.

We further compare RISK-KV against replaying the complete history of tokens during request restoration. Token Replay reruns the historical prefix. RISK-KV uses retained exact KV to reconstruct the omitted cache and avoids that full historical recomputation.

The local experiment uses an RTX 3060 Laptop GPU, a 512-token history, and batch size 1. Each model contributes eight validation requests. After five warm-up iterations, each method is measured 30 times per request. The suspended state resides in local CPU memory. Token Replay reads token IDs and performs one prefill over the 512 historical tokens plus the known current token. RISK-KV transfers retained exact KV for the 512 historical positions, builds two-sided anchor features, reconstructs omitted layers with a Folded FP32 predictor already resident on the GPU, assembles the full cache, and processes that same known current token. Both methods therefore produce next-token logits at the same position. Timing includes CPU-to-GPU state transfer, cache recovery, and the next inference step; the model and predictor are loaded and warmed before measurement. The complete GPU cache is present during subsequent decoding.

\begin{figure}[t]
    \centering
    \includegraphics[width=\linewidth]{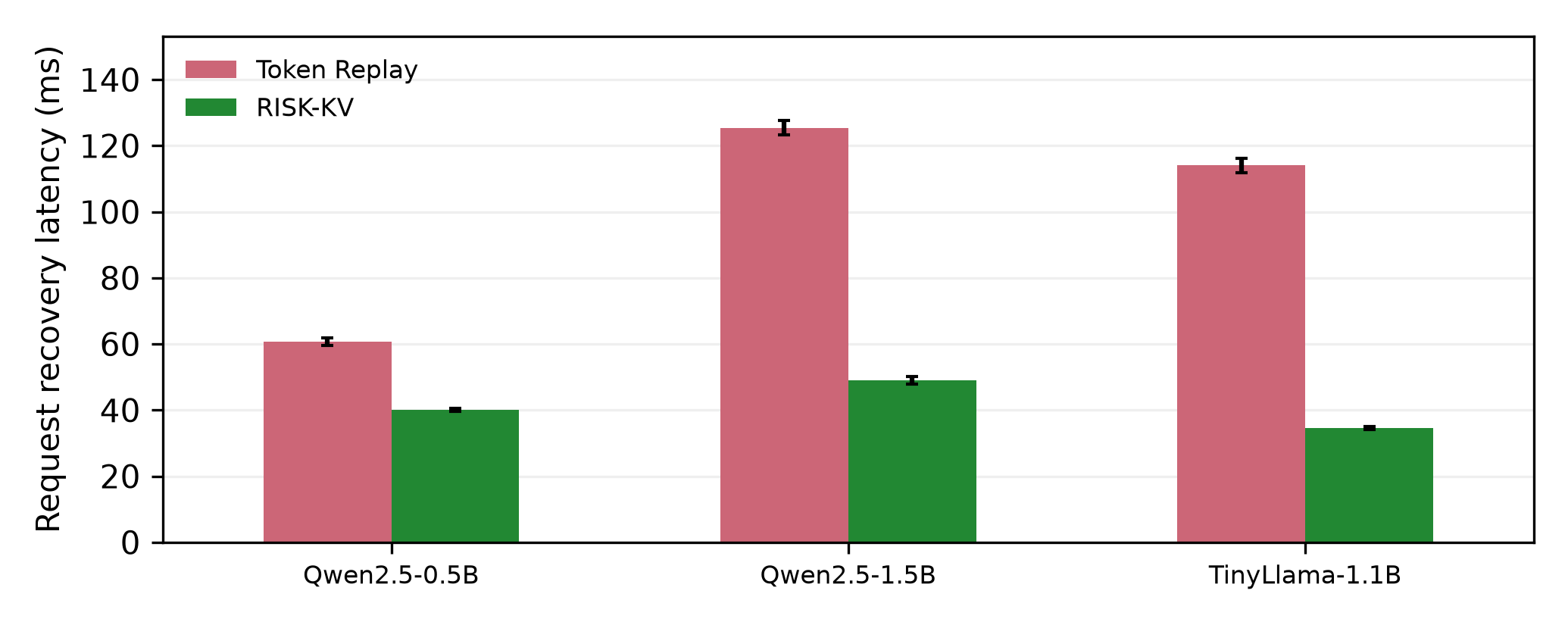}
    \caption{Request-recovery latency of Token Replay and RISK-KV with a 512-token history. Error bars are sample standard deviations across eight requests per model and 30 repetitions per request.}
    \label{fig:request_recovery_latency}
\end{figure}

As Fig.~\ref{fig:request_recovery_latency} shows, RISK-KV has lower recovery latency for all three models. For Qwen2.5-0.5B, Qwen2.5-1.5B, and TinyLlama-1.1B, Token Replay averages 60.66, 125.41, and 114.07 ms, whereas RISK-KV averages 40.09, 48.99, and 34.50 ms. The corresponding speedups are 1.51$\times$, 2.56$\times$, and 3.31$\times$. With a resident predictor, reconstructing historical KV from a partial exact cache reduces the work otherwise spent replaying the prefix.

The selected recovery groups omit 33.3\%, 28.6\%, and 18.2\% of raw historical KV for the three models, respectively. Requests using the same model share the Folded FP32 predictor weights.

\section{Conclusion}
We proposed RISK-KV, a quality-aware method for selective KV-cache reconstruction in distributed LLM inference. It uses exact KV retained on both sides of a target group to reconstruct missing historical KV, assesses functional risk through predictive loss caused by actual reconstruction, and selects groups under a quality budget. Folding the learned recovery map with the model's original KV projections reduces predictor storage without changing pretrained model weights or replaying the complete historical prefix.

Across three pretrained language models, layer-specific reconstruction risks vary substantially. Risk-aware selective recovery reduces historical KV storage by 18.2\%--33.3\% in representative configurations with small observed changes in predictive quality. Conservative quality gates reduce observed budget violations in an independent evaluation sample. For local restoration with a 512-token history, RISK-KV is 1.51$\times$--3.31$\times$ faster than complete Token Replay, demonstrating potential to reduce historical-state recomputation. Actual wireless cross-node migration has not yet been evaluated. Future work will examine longer contexts, larger models, and real distributed deployments.

\bibliographystyle{IEEEtranN}
\bibliography{references}

\end{document}